\documentclass{emulateapj}
\usepackage{subeqn}
\usepackage{verbatim}

\shorttitle{The Atomic Gas PDF of Perseus}

\shortauthors{BURKHART et al.}
\begin{document}

\title{The Lognormal Probability Distribution Function of the Perseus Molecular Cloud: A Comparison of HI and Dust}

\author{ Blakesley Burkhart\altaffilmark{1}, Min-Young Lee\altaffilmark{2}, Claire E. Murray\altaffilmark{3}, Snezana Stanimirovic\altaffilmark{3} }
\altaffiltext{1}{Harvard-Smithsonian Center for Astrophysics, 60 Garden St., Cambridge, MA 0213}
\altaffiltext{2}{Laboratoire AIM, CEA/IRFU/Service d' Astrophysique, Bat 709, 91191 Gif-sur-Yvette, France}
\altaffiltext{3}{Department of Astronomy, University of Wisconsin, Madison, WI 53706}

\begin{abstract}
The shape of the probability distribution function (PDF) of molecular
clouds is an important ingredient for modern theories of star formation
and turbulence. Recently, several studies have pointed out observational difficulties with constraining
the low column density (i.e. A$_V <1$) PDF using dust tracers.   In order to constrain the shape and properties of the 
low column density probability distribution function, we
investigate the PDF of multiphase atomic gas in the Perseus molecular cloud using opacity-corrected GALFA-HI data and compare the PDF shape and properties to the total gas PDF and the N(H$_2$) PDF.  We find that the
shape of the PDF in the atomic medium of Perseus is well described by a
lognormal distribution, and not by a power-law or bimodal
distribution.   The peak of the
atomic gas PDF in and around Perseus lies at the HI-H$_2$ transition column density for
this cloud, past which the N(H$_2$) PDF takes on a powerlaw form. We find that the PDF of the atomic gas is narrow and at  column densities larger than the HI-H$_2$ transition the HI rapidly depletes, suggesting that the HI PDF may be used to find the HI-H$_2$ transition column density.
We also calculate the sonic Mach number
of the atomic gas by using HI absorption line data, which yields a median
value of $M_s=4.0$ for the CNM, while the HI emission PDF, which traces both the WNM and CNM,  has a width more consistent with transonic turbulence.

\end{abstract}

\section{Introduction}
\label{intro} 

Understanding the process of star formation is complicated by the interaction of gravity, magnetic fields, stellar feedback, and
turbulence.  
The details of
the collapse of molecular clouds on parsec scales may determine the key properties of the initial mass function and the 
star formation rates in galaxies (e.g. see Hennebelle \& Falgarone 2012).
Thus in order to obtain a  complete picture of the process of star formation and the evolution of galaxies, the dynamics of
molecular clouds must be understood.

Towards this end, the column density probability distribution function (PDF) is a common tool used by theorists and observers alike to disentangle the role of turbulent motions, stellar feedback, gravitational contraction and magnetic support. 
Several models claim that the high density end of the PDF should
provide the dense gas fraction (Krumholz \& Mckee 2005; Hennebelle \& Chabrier 2011; Padoan \& Nordlund 2011). Furthermore, the highest column density regime of the PDF of self-gravitating turbulent clouds has a power-law distribution as demonstrated in numerical simulations (Federrath \& Klessen 2012, 2013; Collins et al. 2012; Burkhart, Collins \& Lazarian 2015)  and observations (Kainulainen et al. 2009;  Froebrich \& Rowles 2010; Schneider et al. 2013, 2105; Lombardi, Alves \& Lada 2015).

The lower column density material in the PDF of molecular clouds should trace primarily
non-self-gravitating gas, be dominated by turbulent motions, and is expected to have a
lognormal form (Vazquez-Semadeni 1994; Padoan et al. 1997;  Burkhart \& Lazarian 2012).
Lognormal PDFs have been observed in multiple phases in the ISM including CO (Brunt 2010), in dust (e.g., Kainulainen \& Tan 2013; Froebrich \& Rowles 2010; Schneider et al. 2013; Alves de Oliveira et al. 2014), 
and in the diffuse ISM (Berkhuijsen \& Fletcher 2008; Burkhart et al. 2010; Burkhart, Gaensler, Lazarian 2012; Iacobelli et al. 2014). 
The PDF has also been noted by several authors to be an an important method for determining the sonic Mach number  (${\cal M}_s$, Federrath et al. 2008; Burkhart et al. 2009; Kainulainen \& Tan 2013; Burkhart \& Lazarian 2012).  However
Burkhart, Collins, \& Lazarian (2015) have reported that the PDF-sonic Mach number relation should only be robust for diffuse gas
and not in a self-gravitating medium.

Tracing the PDF from observations is non-trivial as multiple tracers must be used in order to sample the largest dynamic range of densities.  
For dense material in molecular clouds, dust emission and absorption are the most promising tracers of the true column density distribution with the largest dynamic range of densities. This is in contrast to molecular line tracers such as CO, which suffer from opacity, excitation temperature, absorption and depletion effects (Goodman et al. 2009; Burkhart et al. 2013a,b).
However, recently several authors (e.g. Schneider et al. 2015; Lombardi, Alves, \& Lada 2015) have noted that dust emission and extinction are problematic probes of the low column density material in molecular clouds.  The PDF as measured from observations of dust can suffer several biases including resolution, noise, boundary effects and  line-of-sight contamination from the foreground and background. Particularly, Lombardi, Alves  \& Lada (2015)
pointed out that the lognormal portion of the PDF (i.e. the portion of the PDF traced below A$_k < 0.1$ mag or A$_V < 1$ mag) may not be securely traced by dust.  In light of this, other tracers should be employed to determine the properties of the PDF in molecular clouds at lower densities.

Our motivation for this Letter is to explore the use of HI as a tracer of the low column density PDF in star forming 
molecular clouds. In general, HI is an excellent tracer of low column density atomic gas in the far outskirts of molecular clouds. While it can be optically thick closer to GMC centers (see Bihr et al. 2015), HI absorption measurements can be obtained to provide full column density information and provides insight into the nature of the atomic to molecular transition in GMCs (Lee et al. 2015).  In particular we focus our study on the Perseus molecular cloud, whose atomic and molecular gas and dust content, as well as the HI-H$_2$ transition,  has been well-studied in a number of previous works (Lee et al. 2012; Lee et al. 2014; Stanimirovic et al. 2014; Lee et al. 2015).

This Letter is organized as follows: in Section 2 we provide a description of the HI and dust data sets of the Perseus molecular cloud  used in this study. In Section 3 we discuss our findings regarding the PDF of HI in and around Perseus and compare it to the dust PDF.  In Section 4 we discuss our results followed by our conclusions in Section 5.

\section{Data}

In this section, we provide a summary of the data in our study and refer to Lee et al. (2012; 2015) and Stanimirovi\'c et al. (2014) for more details. 
\label{data} 
\subsection{HI Emission and Absorption Data}

We use the HI data cube from the GALFA-HI Survey (Peek et al. 2011), 
centered at (R.A.,decl.) = (03$^{\rm h}$29$^{\rm m}$52$^{\rm s}$,$+$30$^{\rm \circ}$34$'$1$''$) in J2000
with a size of $\sim$15$^{\circ}$ $\times$ 9$^{\circ}$. 
To derive the HI column density, Lee et al. (2012) integrated the HI emission 
from $V_{\rm LSR}$ = $-$5 km s$^{-1}$ to $+$15 km s$^{-1}$ under the optically thin assumption. 
This HI velocity range for Perseus was determined based on the maximum correlation between 
the derived $N$(HI) and the 2MASS $A_{V}$ from the COMPLETE Survey (Ridge et al. 2006). 
To construct an accurate PDF of atomic gas, we use the opacity-corrected HI column density image (see Lee et al. 2015 for details), 
but exclude the regions with $N$(HI) $< 6 \times$ 10$^{20}$ cm$^{-2}$ from our analysis 
due to possible contamination with the Taurus molecular cloud (see Figure 1). 

We also use HI absorption data from Stanimirovi\'c et al. (2014) in this work to estimate the sonic Mach number in the CNM of Perseus. 
The HI emission-absorption spectral pairs were obtained toward 26 radio continuum sources located behind Perseus,  
and the Gaussian decomposition method (Heiles \& Troland 2003a) was employed to calculate
the properties of individual CNM and WNM components 
(e.g. spin temperature, optical depth, ``true'' HI column density, etc.).

\subsection{$A_{V}$ based on Far-infrared Dust Emission Data}

We use the total gas (hereafter referred to as the $A_{V}$ map) and N(H$_2$) images from Lee et al. (2015) and refer to that work for details on the data processing. 
The N(H$_2$) image is created from the $N$(HI) and A$_V$ images as: 
\begin{equation}
N(H_2)=\frac{1}{2}(\frac{A_V}{D/G}-N(HI))
\label{eq:subt}
\end{equation}
using the local dust-to-gas ratio (D/G) of $\sim$1 $\times$ 10$^{-21}$ mag cm$^{2}$  from Lee et al. (2015). 
 This is D/G value we adopt for the comparison between the HI and dust PDFs (Section 3). 
In this case, the N(H$_2$) PDF represents the molecular gas without the atomic contribution.
The final $A_{V}$ image (Figure 1) is at 4.3$'$ angular resolution with a pixel size of 4.3$'$ 
(same for the $N$(HI) image),  and the median 1$\sigma$ uncertainty in $A_{V}$ is $\sim$0.17 mag. 

\begin{figure*}[h!]
\begin{center}
\includegraphics[width=0.98\columnwidth]{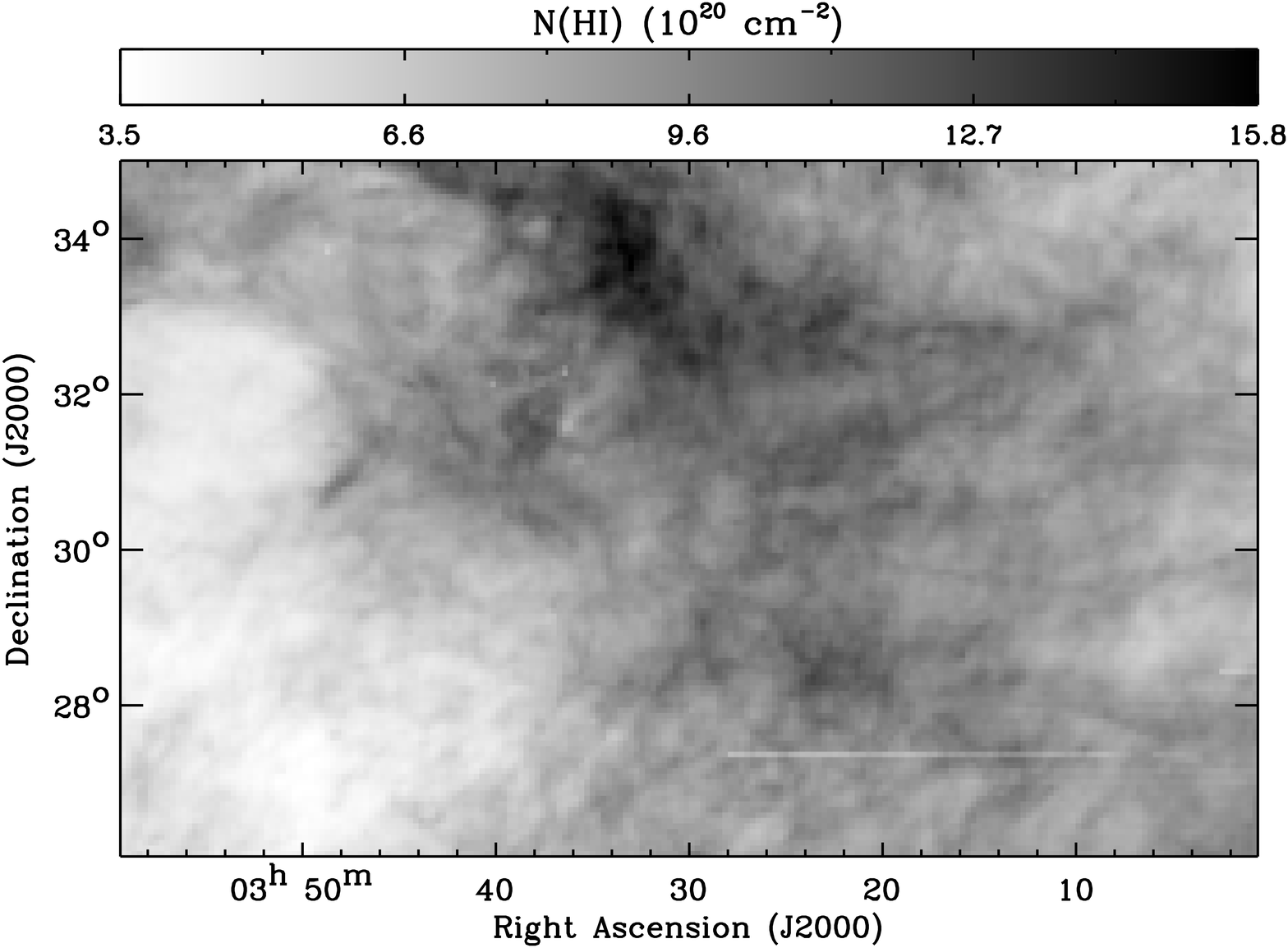}
\includegraphics[width=0.98\columnwidth]{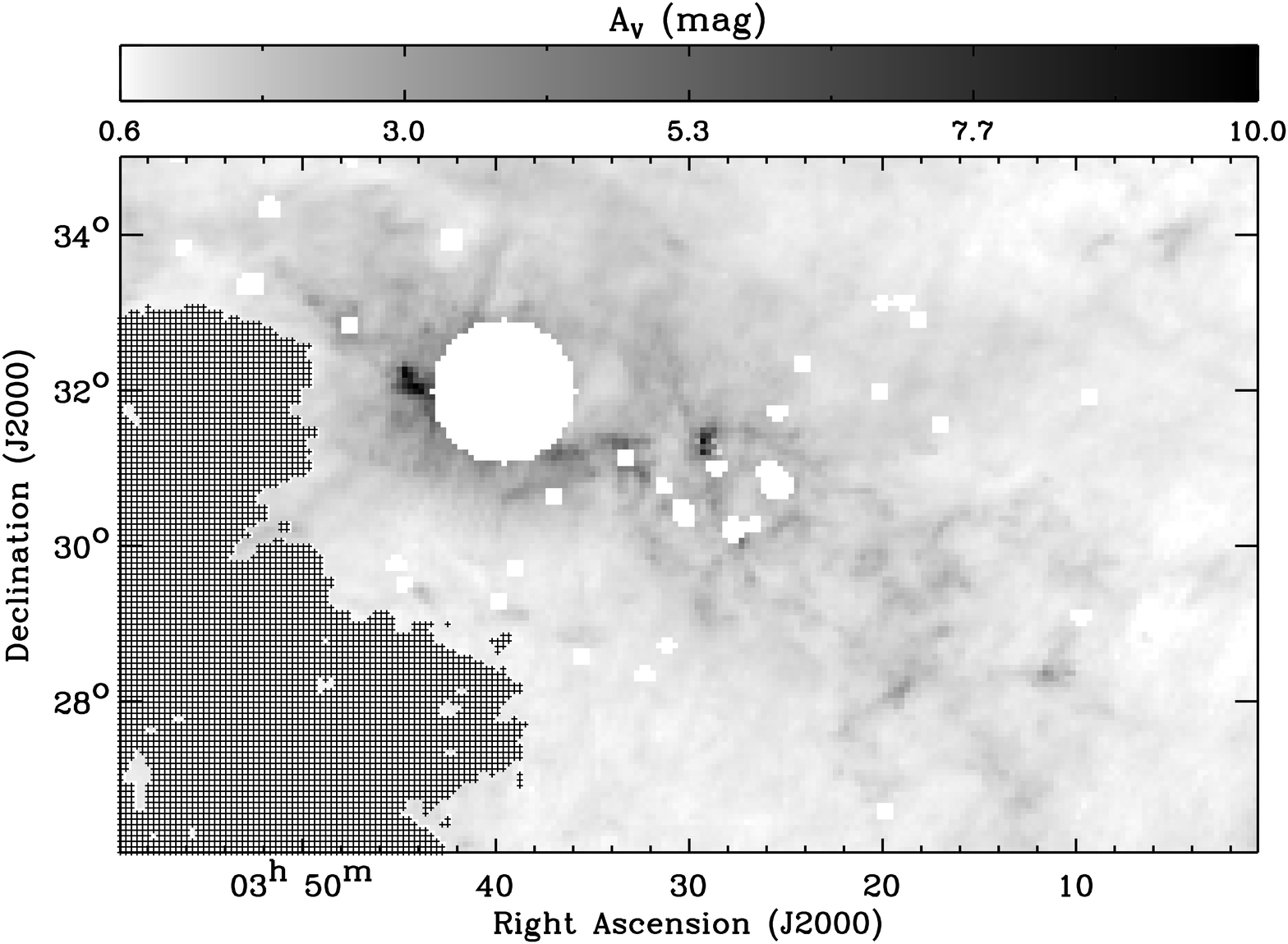}

\caption{Top: Opacity-corrected HI column density image from Lee et al. (2015). Bottom: A$_V$ image from Lee et al. (2015). The blank pixels correspond to the regions with possible contamination, e.g. point sources and a background HII region. In addition, the regions with N(HI) $< 6 \times 10^{20}$ cm$^{-2}$ are shown with black crosses and are not used for our analysis as these pixels could be associated with the Taurus molecular cloud. See Lee et al. (2012) for details on the pixel masking.}
\end{center}
\end{figure*}

\section{Results}
\label{pdf} 

\begin{figure*}[h!]
\begin{center}
\includegraphics[scale=0.7]{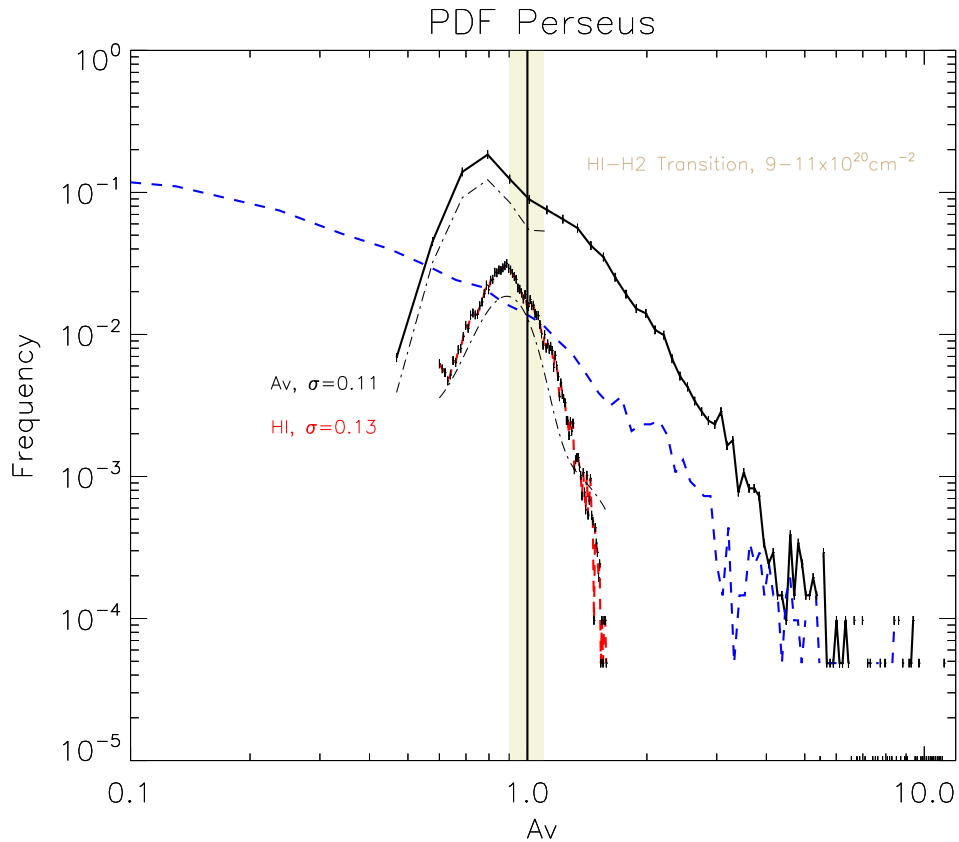}
\caption{The PDF of atomic and molecular gas in the Perseus cloud using the $A_{V}$ and N(H$_2$) images from Lee et al. 2015 (black and blue dashed lines, respectively) and the GALFA 21cm emission (red dashed line), as discussed in the text. The dotted black line directly under the red line is a Gaussian fit to the HI PDF, downshifted for ease of visibility. The PDFs are normalized by the total number of data points, which are identical over both data sets.  The tan box brackets the range of densities for the HI-H$_2$ transition as derived in Lee et al. (2015). The solid black vertical line is a references for $A_V=1$).
The median 1-$\sigma$ error level for the HI data is 0.08 mag (when D/G = $1\times 10^{-21}$ is adopted) and for the A$_V$ data 0.17 mag.}
\end{center}
\end{figure*}

We show the PDF of atomic and molecular gas associated with the Perseus molecular cloud in Figure 2. The black line shows the  PDF of the $A_{V}$ image from Lee et al. (2015), which peaks close to A$_V$ = 1 mag (straight vertical line), which is the column density threshold Lee+15 found for the HI-to-H2 transition in Perseus (tan box), as well as the A$_V$ limit Lombari, Alves \& Lada (2015) suggested for dust as a unreliable tracer of the low column density PDF. The red dashed line shows the PDF of atomic gas.   The blue line shows the N(H$_2$) PDF from Lee et al. (2015).
The PDFs as shown in Figure 2 were binned into 100 bins and have been normalized by the total number of points in the images  (which are the same in both the HI image and the dust images).   The skewness and kurtosis of the  HI GALFA data is  log(HI), 0.1 and -0.05 respectively, which suggest the distribution is nearly Gaussian and is within the range expected from MHD simulations with no gravity (i.e. see Burkhart, Collins \& Lazarian 2015, their Figure 6).  We fit a Gaussian distribution to the HI PDF and the A$_V$  (black lines  downshifted for ease of comparison with the shape)  and find the fitted standard deviation to be $\sigma_{HI}=0.13$ and $\sigma_{A_V}=0.11$.   The 1-sigma errors in the HI image are 0.08 mag (when D/G = $1\times 10^{-21}$ is adopted). We overplot Poisson error bars for both the A$_V$ and HI PDFs.

The HI gas associated with Perseus can be well described by a narrow lognormal distribution which peaks close to $A_{V}=1$ mag, which we mark with a straight vertical line.  The shape of the N(H$_2$) PDF at column densities above the HI-H$_2$ threshold is dominated by a power law tail, similar to the A$_V$ PDF, while at low column densities below the HI-H$_2$ threshold the N(H$_2$) PDF flattens due to the fact that there is a large amount of HI-dominated diffuse gas around Perseus, which is subtracted from the low amount of total gas via Equation 1. The shape of the total gas A$_V$ PDF (which includes both N(H$_2$) and HI gas) below the HI-H$_2$ threshold is dominated by a lognormal shape (i.e. the HI PDF) while above it displays a powerlaw tail.  The fact that the N(H$_2$) PDF is not lognormal at column densities below the  HI-H$_2$ threshold suggests that the lognormal portion of the A$_V$ PDF is produced by the HI gas, which of course is removed to create the N(H$_2$) PDF. The lognormal behavior of the atomic gas is in agreement with numerical simulations of supersonic MHD turbulence (e.g. Burkhart \& Lazarian 2012; Federrath \& Klessen 2012; Burkhart Collins \& Lazarian 2015) and only the dense self-gravitating molecular gas displays a power-law tail behavior.

We note that the PDF of the HI column density distribution for Perseus is narrow, which is expected considering the very uniform distribution of HI in this molecular cloud (see Lee et al. 2012; 2015).
This is because hydrogen is converted from atomic to molecular (HI-to-H$_{2}$ transition) 
at $N$(HI) $\sim$ (9--11) $\times$ 10$^{20}$ cm$^{-2}$ for Perseus (Lee et al. 2015).  
However, in general the uniform HI column density distribution could also result from a large amount of optically thick HI gas (see Bihr et al. 2015). The amount of the optically thick HI in Perseus was studied in Lee et al. (2015).  They found that the contribution from the optically thick HI is minor (less than 20\%) and H$_2$ formation is indeed responsible for the nearly uniform HI column density across Perseus (see also Bialy et al. 2015).  The fact that the HI distribution is not bimodal\footnote{Bimodal distributions of HI are also predicted due to the fact that this tracer probes a multiphase medium with regimes of thermal instability, see Gazol \& Kim (2013).} suggests that much of the gas may be close the to transition to molecular gas.

Interestingly, the HI and A$_V$ PDFs peak at comparable column density values (within the 1-$\sigma$ errors of the data) and have similar values for the fitted Gaussian widths, however a number of effects must also be considered. 
For the HI PDF, the peak value will depend on the D/G that is used as well as the cloud superpositions along the line of sight, as well as the amount of cold HI.   These effects have been well-studied in Perseus and  the line-of-sight velocity which is used to disentangle different clouds, as well as correction for high optical depth, is well constrained (see Lee et al. 2015).
For the dust PDF, the peak location depends on the LOS contamination, offset corrections  and superposition effects (see Schneider et al. 2015; Lombardi, Alves \& Lada 2015). 
Stanchev et al. (2015) demonstrated different selected regions in the Perseus PDF of \textit{Planck} dust optical depth at 353 GHz ($\tau_{353}$) would show different values of the PDF peak and width.
This is most likely the reason why the peak of the A$_V$ PDF for Perseus in the Lee et al. (2015) data is offset from the peaks of the Kainulainen et al. (2009) and Sadavoy et al. (2014) Perseus PDFs (A$_V \approx 0.8$ as compared to A$_V \approx 1.5$, respectively).
The width of the Perseus PDF reported in Kainulainen et al. (2009) is much larger ($\sigma=0.47$) than the width found in our work, which is more compatible to the value reported by  Stanchev et al. (2015) using Planck data, i.e. $\sigma \approx 0.11$.

\subsection{The Sonic Mach Number}

Numerical simulations of MHD turbulence have suggested that the sonic Mach number is related to the width of the lognormal PDF (e.g. see  Federrath et al. 2008, Burkhart \& Lazarian 2012 and references therein).  We compute the sonic Mach of the HI gas of Perseus using both the PDF and CNM absorption measurements.

First, we compute the sonic Mach number using the absorption line data of Stanimirovic et al. (2014)  and  Equation 17 of Heiles \& Troland (2003):
\begin{equation}
M_s^2=4.2(\frac{T_{k,max}}{T_s}-1)
\end{equation}
\label{fig:mach1} 

We show a histogram of the Mach numbers derived from Equation 1 using 107 CNM absorption components (described in Section 2 and in Stanimirovic et al. 2014) in Figure 3.  The median value is $M_s=4.0$, and is indicated by the vertical dashed line.  

\begin{figure}[h!]
\begin{center}
\includegraphics[width=0.97\columnwidth]{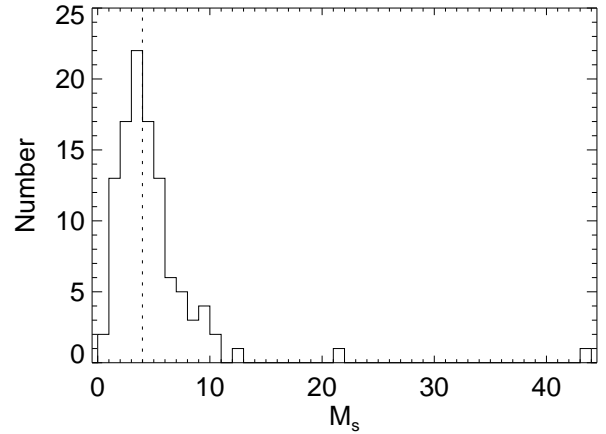}
\caption{Histogram of the sonic Mach number as calculated from the absorption line data for Perseus. The median value is $M_s=4.0$ and is shown with a straight vertical line. }
\end{center}
\end{figure}

Since the HI distribution shows a single lognormal distribution we can estimate the sonic Mach number from the width of the lognormal using Equation 4 from Burkhart \& Lazarian (2012): 

\begin{equation}
 \sigma_{ln \Sigma/\Sigma_0}^2=A\times ln(b^2{\cal M}_s^2+1)
 \end{equation}
 
for different values of parameters $A$ (which depends on the optical depth, see Burkhart et al. 2013a) and $b$ (which depends on the type of turbulence driving, see Federrath et al. 2008).  Figure 4 shows curves of Equation 2 for different values of $A$ and $b$. The narrow PDF of the HI gas  in Perseus is consistent with transsonic Mach numbers or with very low values of the parameter $A$.  Low values of $A$ were shown to be consistent with a more optically thick medium (Burkhart et al. 2013a), therefore it is more likely that the PDF of the HI gas, which is tracing a mix of the WNM and CNM, has a lower average sonic Mach number than the sonic Mach number reported from absorption lines, which only trace the CNM.

\begin{figure}[h!]
\begin{center}
\includegraphics[width=0.97\columnwidth]{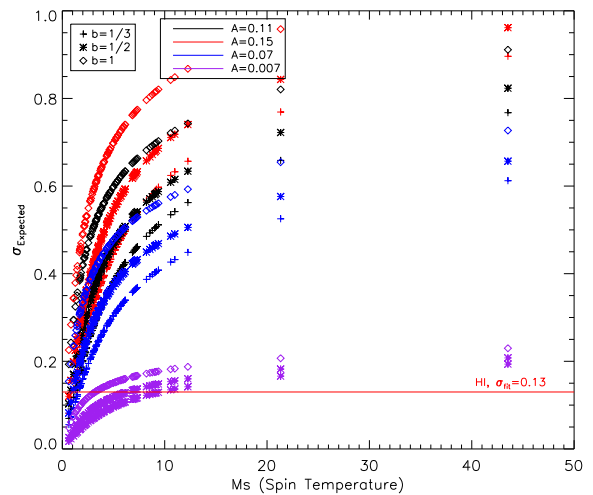}
\caption{Variance- sonic Mach number relationship given by Equation 2 for different values of $A$ and $b$. We use the same distribution of sonic Mach number as presented in Figure 3. }
\end{center}
\end{figure}

\section{Discussion}
\label{dis} 

The shape and properties of the PDF of molecular clouds are important ingredients for modern theories of star formation.
In this letter we found that the low column density material in the Perseus molecular cloud as traced by neutral hydrogen takes on a lognormal shape rather than a bimodal or power-law distribution.  A number of processes can induce  a lognormal distribution in density, including subsonic and supersonic turbulence  but also gravitationally contracting clouds which are not necessarily turbulent (see Tassis et al. 2010).   
However, high Reynolds number flows (with Re $=LV/\nu > 10^3$) are almost always turbulent\footnote{The Reynolds number in molecular clouds can be upwards of Re $\approx 10^8$, assuming 
a typical GMC viscosity of $\nu \approx 10^{16} cm^2 s^{-1}$, cloud scale $L$=4pc and velocity  $V$=1km/s. Larger values of $L$ or $V$ will only increase the value of Re.}  and the density of HI is diffuse enough to ignore the effects of self-gravity.
If turbulence is indeed responsible for the lognormal HI PDF, 
our study along with Lee et al. (2012, 2015) and Bialy et al. (2015) suggests that 
the chemical and dynamical evolution of the interstellar medium can be modeled separately, at least for low-mass molecular clouds like Perseus.  In Lee et al. (2012, 2015) and Bialy et al. (2015), steady state and chemical equilibrium models of H$_{2}$ formation (Krumholz et al. 2009; Sternberg et al. 2014) 
were tested against the observations of Perseus and a good agreement was found. 
This good agreement implies that, at least for clouds where strong stellar feedback is not important,
H$_{2}$ formation can be modeled in steady state 
while turbulence determines the density distribution of the cloud. 

In this letter we derived the sonic Mach number of the atomic gas in and around Perseus using two independent methods: the width of the PDF of the HI column density and absorption line fitting using data from Stanimirovic et al. (2014).  The median Mach number derived from the absorption line data ($M_s=4.0$).  On the other hand, the width of the HI PDF indicates that HI (CNM + WNM) in and around Perseus is on average transsonic  with the sonic Mach number lower than what the HI absorption line data and CO observations suggest 
(e.g. Padoan et al. 2003; M$_s > 10$ for Perseus based on CO observations).
The 21-SPONGE survey (Begum et al. 2010; Murray et al. 2014, 2015)
will provide additional absorption line data using the VLA  and will be able to further test the HI PDF-Mach number relation for other clouds.  Ultimately knowing the Mach number in the atomic gas will constrain the initial conditions for numerical simulations of diffuse and star forming clouds.

It should be pointed out that there are observational difficulties to derive accurate HI column densities associated with molecular clouds. In particular one must investigate clouds that lie off the plane of the Galaxy to mitigate line-of-sight confusion (e.g. Perseus at the Galactic latitude of $-$20$^{\circ}$). It is also important to define the HI velocity range for an individual molecular cloud, correct for high optical depth HI, and check that the HI correlates to some degree with the dust map. All of these procedures were performed for the data set used in this study (see Lee et al. 2012; 2015).

Despite the difficulties in accurately constraining the PDF of molecular clouds through HI, the atomic gas PDF should be compared to the molecular PDFs (i.e. traced by dust emission/absorption) whenever possible as they probe different column density regimes.  
The peak of both the $A_V$ and HI PDFs explored in this paper lies roughly at the HI-to-H$_2$ transitional column density found in Lee et al. (2012) for Perseus. The HI PDF peak in GMCs may trace this transitional density since the HI rapidly depletes at higher column densities to H$_{2}$.  Future studies  should investigate the HI column density distribution in other GMCs and compare with the measured HI-to-H$_2$ transition as our study suggests the transition between the lognormal to power-law behavior of the PDF traces the HI-to-H$_2$ transition.  

\section{Conclusions}
\label{con} 

We investigate the multiphase PDF of the atomic gas in the Perseus molecular cloud and compare it to the PDF traced by dust.  We find that:

\begin{itemize}

\item The shape of the PDF of the atomic gas in Perseus, which traces gas with $A_V<1$ mag,   is well described by a lognormal distribution, and not by a power-law distribution or a bimodal distribution.   

\item The shape of the N(H$_2$) PDF is a power law past the HI-to-H$_2$ transition, similar to the total gas ($A_V$) PDF.

\item The transition from the lognormal atomic PDF to the power-law dust PDF happens  roughly at the HI-to-H$_2$ transition.
This suggests that the HI and dust PDFs may be used in other clouds to find the HI-to-H$_2$ transition.

\item We  calculate the sonic Mach number
of the atomic gas by using HI absorption line data, which yields a median
value of $M_s=4.0$ for the CNM while the HI emission PDF, which traces both the WNM and CNM,  has a width more consistent with transonic turbulence.

\end{itemize}

B.B. is thankful for valuable discussion with Simon Bihr, David Collins, Alyssa Goodman, Nia Imara, Eric Keto, Charlie Lada, and Phil Myers. The research of B.B. is supported by the NASA Einstein Postdoctoral Fellowship.  M.-Y. Lee is supported by the SYMPATICO grant (ANR-11-BS56-0023) from the French Agence Nationale de la Reserche and the DIM ACAV.  C.E.M acknowledges support from the NSF Graduate Research Fellowship. S.S acknowledges support from the NSF Early Career Development (CAREER) Award AST-1056780.

\end{document}